\documentclass{article}
\usepackage{spconf,amsmath,graphicx}

\usepackage{cite}
\usepackage{txfonts}
\usepackage{setspace}
\usepackage{bm}
\usepackage{multirow}
\usepackage{subfig}
\usepackage{comment}
\usepackage{amssymb}
\usepackage[dvipdfmx]{hyperref}
\usepackage{xcolor}
\usepackage{setspace}
\usepackage{arydshln}

\title{Improving Character Error Rate Is Not Equal to Having Clean Speech:\\
Speech Enhancement for ASR Systems with Black-box Acoustic Models
}

\name{Ryosuke Sawata, Yosuke Kashiwagi and Shusuke Takahashi} 
\address{Sony Group Corporation, Tokyo, Japan}

\begin{document}
\ninept
\maketitle
\begin{abstract}
A deep neural network (DNN)-based speech enhancement (SE) aiming to maximize the performance of an automatic speech recognition (ASR) system is proposed in this paper.
In order to optimize the DNN-based SE model in terms of the character error rate (CER), which is one of the metric to evaluate the ASR system and generally non-differentiable, our method uses two DNNs: one for speech processing and one for mimicking the output CERs derived through an acoustic model (AM).
Then both of DNNs are alternately optimized in the training phase.
Even if the AM is a black-box, e.g., like one provided by a third-party, the proposed method enables the DNN-based SE model to be optimized in terms of the CER since the DNN mimicking the AM is differentiable.
Consequently, it becomes feasible to build CER-centric SE model that has no negative effect, e.g., additional calculation cost and changing network architecture, on the inference phase since our method is merely a training scheme for the existing DNN-based methods.
Experimental results show that our method improved CER by 8.8\% relative derived through a black-box AM although certain noise levels are kept.
\end{abstract}
\vspace{-2mm}
\section{Introduction}
\label{sec:intro}
\vspace{-2mm}
Speech enhancement (SE), which aims to extract target speech from noisy input signal, remains an important research topic in the field of sound source separation.
In particular, many deep neural network (DNN)-based approaches have appeared in recent years and they have been shown to dramatically outperform conventional methods.
For instance, convolutional neural networks (CNNs) \cite{cnn_org} have been shown to be better than using a short-time Fourier transform (STFT) and inverse STFT (ISTFT) for building an encoder and decoder \cite{conv_tasnet}.
Furthermore, methods that utilize recurrent neural networks (RNNs)-based models have been shown to be capable of real-time processing \cite{rnn_se_1, rnn_se_2, rnn_se_3}.
In addition, there are hybrid methods that exploit the benefits of both types of network, i.e., real-time processing and high performance \cite{crnn_hybrid_se_1, crnn_hybrid_se_3}.

The aforementioned methods have dramatically improved the performance of the SE task in terms of some subjective and objective metrics related to the perceptual quality of processed signals.
However, it has been reported that the performance of automatic speech recognition (ASR) is often tended to be degraded by using the signals processed by the DNN-based SE models as input \cite{DNNbased_SingleSE_degradeASR_1, DNNbased_SingleSE_degradeASR_2, DNNbased_SingleSE_degradeASR_3, DNNbased_SingleSE_degradeASR_4, DNNbased_SingleSE_degradeASR_5}.
In other words, the output of DNN-based SE models tends to degrade the quality of the text results recognized by ASR system.
The acoustic model (AM), which is the part of the ASR system that models the relationship between the audio signal and the phonetic units in the language, is typically trained in a data-driven manner, and thus, its performance is strongly affected by mismatches between the training and testing conditions.
In general, the output signals of SE models have various types of signal distortion.
Particularly, the distortion caused by a DNN-based SE model is considered especially serious because a DNN is a powerful and complex non-linear calculator that is rather flexible compared with the conventional models.
Therefore, such various types of signal distortion cause mismatches with training data and consequently degrade the performance of ASR\footnote{The same mismatch is expected even in End-to-End speech recognition, which does not explicitly train AMs.}.

A straightforward way to solve the above problem is to perform additional training of the AM by using signals obtained from the output of a DNN-based SE model, which is called ``re-training''\cite{joint_se_am_3}.
Furthermore, joint training, which connects the DNNs of the SE and AM models and trains the whole connected model as one DNN in terms of the AM's criteria, has been shown to be an effective way to improve the performance of ASR\cite{joint_se_am_1, joint_se_am_2}.
Since the AM becomes able to know the data provided by the DNN-based SE model via re-training or joint training, it is obvious that the performance of ASR system can be improved.  
However, when trying to use a DNN-based SE model to provide training data to the AM, these approaches need a huge amount of data to be processed by the SE model and correspondingly long additional time for training.
In particular, the variations related to the SE task, i.e., signal-to-noise ratio (SNR), reverberation, the kinds of noise and so on, make it very laborious.
Furthermore, in many real-world applications, the ASR system can be supplied by a third party. In such case, it becomes difficult to apply re-training and joint training to the AM.
Therefore, 
it is desirable that SE model alone become able to deal with the mismatch between the data of SE and AM. 
As far as we know, although some studies have started on this purpose, their results have been limited\cite{se_reinforce_1, se_ceg_1, se_ceg_2}.
For example, the unstableness of the method in \cite{se_reinforce_1} limits its applicability  to only one type of noise.
Moreover, the AM in \cite{se_ceg_1, se_ceg_2} has to be partly white-box since intermediate outputs of the AM, i.e., state posterior probabilities, are used to optimize the front SE model.

To release the above limitations, i.e., enable the DNN-based SE model to be connected with the non-trainable and black-box ASR system effectively, we pay attention to the methods aiming to maximize black-box metric, called MetricGAN\cite{metricgan}. 
On the basis of \cite{metricgan}, some researchers have improved a type of the black-box metric which is obtained by inputting the results of the DNN-based SE model to a black-box function\cite{metricgan_expand, metricgan_kawanaka, metricgan_plus}.
However, all of the metrics that have been examined so far are limited to one kind, i.e., perceptual quality.
In other words, as far as we know, there is no research which has aimed to build a new DNN-based SE model based on \cite{metricgan} in order to maximize the performance of ASR system, e.g., ward error rate (WER), character error rate (CER) and so on. 

Motivated by the above background, we newly exploit a training scheme of DNN-based SE that aims to minimize the CER calculated by using the text results obtained from a black-box ASR system.
Note that our method is not only just applying the MetricGAN \cite{metricgan} by using the CER as the target metric to optimize but also containing the other ideas specialized for minimizing the CER.
Namely, our paper has mainly two contributions: i) applying the MetricGAN \cite{metricgan} to the ASR task, which it has never been studied as far as we know, and ii) enhancing its performance by adding our ideas, which are describe in the following Sec.~\textbf{\ref{subsec:proposed_data}}.
Specifically, our method first builds two DNNs based on MetricGAN \cite{metricgan}: one is used for speech processing, the other is used for mimicking the output CERs derived from an ASR system.
Then, in order to specialize for improving the performance of the ASR system, we add two considerations which differ from the original method: (a) how to normalize data and (b) applying data augmentation.
By optimizing the DNN-based SE model monitoring the mimicked CERs with these considerations, it becomes feasible to build a CER-centric SE model.
Note that there is no negative effect on the inference phase since our method is merely a training scheme; thus, it has high practicability and extensibility for the existing DNN-based SE methods.

\begin{figure}[!t]
\begin{minipage}[b]{1.0\linewidth}
  \centering
  \centerline{\includegraphics[width=8.75cm]{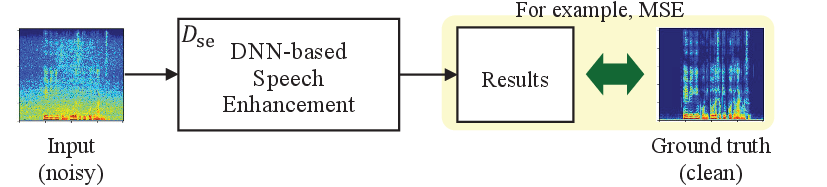}}
  \flushleft{(a) Example of conventional DNN-based SE model}
\end{minipage}
\begin{minipage}[b]{1.0\linewidth}
\vspace{+2.5mm}
  \centering
  \centerline{\includegraphics[width=8.75cm]{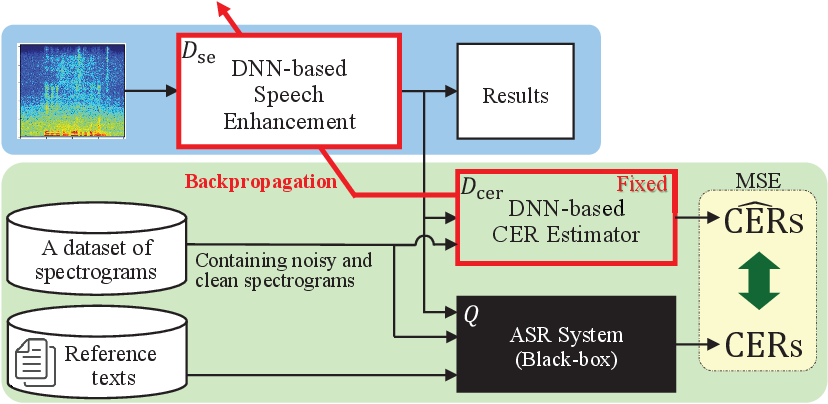}}
  \flushleft{(b) Our CER-centric SE model.
  In the training phase, the light green and blue parts, i.e., the DNN-based CER estimator and the core of SE, are trained alternately.
  Namely, each of them is non-trainable during another part training.}
\end{minipage}
\vspace{-5.5mm}
\caption{Comparison of SE training methods}
\label{fig:overview}
\vspace{-4mm}
\end{figure}
\vspace{-2mm}
\section{Building CER-centric Speech Enhancement}
\label{sec:propose}
\vspace{-2mm}
In this section, we describe our new training scheme for DNN-based SE model, i.e., CER-centric SE model.

\vspace{-2.0mm}
\subsection{Data Preparation}
\label{subsec:proposed_data}
\vspace{-2.0mm}
We focus on our original ideas regarding data preparation, which are different from those of MetricGAN \cite{metricgan}.

\vspace{-2.0mm}
\subsubsection{Normalization}
\label{subsec:data_normalize}
\vspace{-2.0mm}
Here, we describe how to normalize the input and ground-truth spectrograms.
MetricGAN \cite{metricgan} normalizes only the input spectrogram whereas the ground-truth and the output spectrograms are not normalized\footnote{\url{https://github.com/JasonSWFu/MetricGAN}}.
If the spectrograms are not normalized, they can have power bias of the frequency band depending on the type of mixtured noise, the content of utterance, and so on.
This may not be preferable since the frequency bands focused by a black-box ASR system when it recognizes phonemes are diverse, i.e., depending on what AM is used.
Therefore, to enable our following DNN-based CER estimator to deal with any ASR systems effectively, eliminating the power bias depending on the frequency band is preferable in advance.
To do so, we normalize not only the input but also the ground-truth spectrograms.

We assume that the noisy time-domain signal can be expressed as $\bm{x} = \bm{s} + \bm{n}$,
where $\bm{s}$ and $\bm{n}$ are respectively the target and noise signals.
Furthermore, we assume that the DNN-base SE model outputs a mask $\bm{M}$ for extracting the target signal $\bm{\hat{s}}$ as follows:
\begin{align}
\hat{\bm{s}} &= \mathcal{S}^{-1} \{ \hat{\bm{S}} \} = \mathcal{S}^{-1} \{ \bm{M} \circ \bm{X} \},
\label{eq:1}
\end{align} 
where $\mathcal{S}$ and $\mathcal{S}^{-1}$ are the forward and inverse operators of the STFT, respectively.
Here, $\hat{\bm{S}}$ is the predicted spectrum obtained by applying the estimated mask $\bm{M}$ to the input noisy one $\bm{X}$.
Note that $\bm{X}$, $\bm{S}$ and $\bm{M}$ have the same size, i.e., $(F \times N)$, where $F$ and $N$ denote the total numbers of frequency bins and frames, respectively. 
Then, we normalize the mean values of $\bm{X}$ in the horizontal direction:
\begin{align}
\bar{\bm{X}}_{\mbox{\scriptsize{me}}} = \bm{X} \bm{H},
\label{eq:2}
\end{align} 
where $\bm{H} = \bm{I} - \frac{1}{N}\bm{1}\bm{1}^\mathrm{T}$ is a centering matrix, $\bm{I}$ is the $N \times N$ identity matrix, and $\bm{1} = [1, \cdots, 1] \in \mathbb{R}^N$ is an $N$-dimensional vector.
Next, we calculate standard deviation vector $\bm{\sigma} \in \mathbb{R}^F$, each element of which has a corresponding standard deviation in the horizontal direction of $\bm{X}$:
\begin{align}
\bm{\sigma} = \mbox{diag} \left( \sqrt{ \frac{1}{N} \bm{\bar{\bm{X}}_{\mbox{\scriptsize{me}}}} \bm{\bar{\bm{X}}_{\mbox{\scriptsize{me}}}}^\mathrm{T}} \right),
\label{eq:3}
\end{align} 
where the operator `$\mbox{diag}( \bullet )$' extracts the diagonal elements of the input and vectorizes them.
Then, we obtain the normalized spectrum $\bar{\bm{X}}$, each row of which has zero mean and one standard deviation, as follows:
\begin{align}
\bm{\bar{X}} = \bar{\bm{X}}_{\mbox{\scriptsize{me}}} \oslash \bm{\sigma} \bm{1}^\mathrm{T},
\label{eq:4}
\end{align} 
where `$\oslash$' indicates Hadmard division.
MetricGAN employed the above $\bm{\bar{X}}$ as input and applied output mask to the original $\bm{X}$ to estimate the clean spectrum $\bm{\hat{S}}$.
On the other hand, we apply the output mask $\bm{M}$ to $\bar{\bm{X}}_{\mbox{\scriptsize{std}}}$, which is normalized only in terms of the standard deviation; then, we can derive the corresponding $\bar{\bm{S}}_{\mbox{\scriptsize{std}}}$ as follows:
\begin{align}
\bar{\bm{X}}_{\mbox{\scriptsize{std}}} &= \bm{X} \oslash \bm{\sigma} \bm{1}^\mathrm{T}, \\
\bar{\bm{S}}_{\mbox{\scriptsize{std}}} &= \bm{S} \oslash \bm{\sigma} \bm{1}^\mathrm{T}.
\label{eq:5-6}
\end{align}
This means that our input vectors, i.e, every rows of $\bm{X}$ and $\bm{S}$, are normalized in terms of the standard deviation only.
Namely, each element of our input vectors has unit variance, but not mean of zero.

By using these normalized spectrograms, it is expected to exclude the negative effect caused by the power bias depending on the frequency band.

\vspace{-2.0mm}
\subsubsection{Data Augmentation}
\label{subsec:data_specaug}
\vspace{-2.0mm}
In order to assess what type of spectrogram is important for the black-box AM when recognizing phonemes, various types should be input to the ASR system.
In fact, Kawanaka \textit{et al.} reported a performance improvement over their original method when they increased the input types with additional noisy spectrograms\cite{metricgan_kawanaka}.

Inspired by the aforementioned research, we employ a part of SpecAugmentation \cite{specaug}: the time and frequency masking.
Although it is very simple and easy to use because it merely masks a block of consecutive time steps or frequency bands, it has been reported to have beneficial effect on ASR performance.
This is because time and frequency masking probably encourages the ASR system to consider which blocks are important for recognizing phonemes in the input spectrogram.
Thus, we expect that this augmentation will encourage our method to assess the target black-box ASR system, i.e., in which the blocks are important to mimic CER.

\vspace{-2.0mm}
\subsection{DNN-based CER Estimator}
\label{subsec:proposed_wer}
\vspace{-2.0mm}
Now let us describe the DNN-based CER estimator which is trained alternately with the CER-centric SE model explained in the following subsection.
The light green part of Fig.~\ref{fig:overview}(b) is an overview of the procedure to train the CER estimator.

Main purpose of this estimator is searching and understanding the latent feature space of the black-box AM and reflecting it to the DNN.
Then, it is expected to use the estimator as a differentiable function which can calculate CER.
Specifically, we train the DNN by utilizing the following loss function:
\begin{align}
\begin{split}
    \mathcal{L}_{\mbox{\tiny{CER}}} 
        = &\mathbb{E}_{\bm{x}, \bm{s}, t} [ \{D_{\mbox{\tiny{cer}}}(\bm{x}, \bm{s}) 
            - Q(\bm{x}, t) \}^2 \\ 
        &+ \{D_{\mbox{\tiny{cer}}}(\bm{s}, \bm{s}) 
            - Q(\bm{s}, t) \}^2 \\ 
        &+ \{D_{\mbox{\tiny{cer}}}(D_{\mbox{\tiny{se}}}(\bm{x}), \bm{s}) 
            - Q(D_{\mbox{\tiny{se}}}(\bm{x}), t) \}^2 ],
\end{split}
\label{eq:7}
\end{align} 
where $D_{\mbox{\tiny{cer}}}$ and $D_{\mbox{\tiny{se}}}$ are respectively the target DNNs, i.e., the DNN-based CER estimator and SE model.
Note that $D_{\mbox{\tiny{se}}}$ is parameter fixed when minimizing Eq.~\eqref{eq:7}; namely, only the DNN-based CER estimator $D_{\mbox{\tiny{cer}}}$ is trained.
$Q(\bullet, \bullet)$ is a function that calculates the CER by utilizing the target signal and reference text.
Specifically, it has two steps: recognizing the spoken text via the ASR system and calculating the CER by comparing the recognized results with a reference text denoted as `$t$' in Eq.~\eqref{eq:7}.
In this way, the CERs are calculated via the function $Q(\bullet, \bullet)$ as follows:
\begin{align}
    Q(x, t) = \frac{1}{|t|}(\mbox{I}(x,t) + \mbox{D}(x,t) + \mbox{S}(x,t)) \times 100,
    \label{eq:cer}
\end{align}
where $|t|$ indicates the number of the characters in the target text, and $\mbox{I}(x, t)$, $\mbox{D}(x, t)$ and $\mbox{S}(x, t)$ denote the number of insertions, deletions and substitution errors, respectively.
To count these errors, the inferred text and the target text must be aligned, and thus we use a dynamic programming matching technique.
Therefore, this evaluation metric is equivalent to the normalized edit distance.
Note that Eq.~\eqref{eq:cer} can exceed 100\%, especially if there are many insertion errors.
Since we found that large errors have a negative effect on the training, we limit the maximum to 100\% as follows:
\begin{align}
    Q(x, t) = \min\left(\frac{1}{|t|}(\mbox{I}(x,t) + \mbox{D}(x,t) + \mbox{S}(x,t)) \times 100, 100\right).
    \label{eq:cermax}
\end{align}

In this way, by optimizing $\mathcal{L}_{\mbox{\tiny{CER}}}$, the DNN $D_{\mbox{\tiny{cer}}}$ becomes able to mimic the CER calculated through a black-box ASR system.

\vspace{-1.5mm}
\subsection{DNN-based CER-centric Speech Enhancement}
\label{subsec:proposed_se}
\vspace{-1.5mm}
The DNN-based CER-centric SE model is optimized using the estimated CER described in the previous subsection \textbf{\ref{subsec:proposed_wer}}.
The light blue part of Fig.~\ref{fig:overview}(b) is an overview of the training procedure of the SE model.

By utilizing the trained DNN-based CER estimator, the objective function for optimizing CER-centric SE model is defined as follows:
\begin{align}
\mathcal{L}_{\mbox{\tiny{SE}}} 
    = &\mathbb{E}_{\bm{x}} [ \{D_{\mbox{\tiny{cer}}}
            (
                D_{\mbox{\tiny{se}}}(\bm{x}), \bm{s}
            )
        - 0.0 \}^2 ].
\label{eq:10}
\end{align} 
Equation \eqref{eq:10} intends to make the mimicked score of CER close to the ideal score, i.e., zero.
Namely, it is a minimization of the mimicked score of CER derieved by inputting the processed signal $D_{\mbox{\tiny{se}}}(\bm{x})$ to the DNN-based CER estimator $D_{\mbox{\tiny{cer}}}$.
Note that $D_{\mbox{\tiny{cer}}}$ is parameter fixed during this optimization.
In this way, it becomes feasible to build a DNN-based CER-centric SE model.

\vspace{-2mm}
\section{Experiments}
\label{sec:exp}
\vspace{-2mm}
We examined the validity of our method by conducting speech enhancement experiments.

\vspace{-3mm}
\subsection{Setup}
\label{subsec:setup}
\vspace{-1.5mm}
\subsubsection{Simulation of Datasets}
\label{subsec:exp_data}
\vspace{-1.5mm}
In our experiments, we used the following three datasets in order to simulate the training and evaluation sets of noisy speeches.
\\
\\
\textbf{Libri-light \cite{libri_light}} \\
To simulate the set of noisy speeches, we utilized Libli-light \cite{libri_light}.
This dataset is a subset of the original LibriSpeech \cite{libri_org} which is well-known dataset.
Libli-light contains two part, a large part consisting of 60K hours and a small part consisting of about 10 hours in English.
We utilized the small part in our experiments, specifically, 9h worth (2477 utterances) for training and the remaining 1h (286 utterances) for inference and evaluation.
This division between the training and evaluation sets, i.e., into 9h and 1h sets, is given in \cite{libri_light} and its code\footnote{\url{https://github.com/facebookresearch/libri-light}}.
\\
\\
\textbf{MUSAN \cite{musan}} \\
As additive noise in the training, we drew samples from the MUSAN noise dataset \cite{musan}.
This dataset contains various types of noise and has 930 noise files in total.
\\
\\
\textbf{DEMAND \cite{demand}} \\
As additive noise in the inference and evaluation, we drew samples from the diverse environments multi-channel acoustic noise database (DEMAND) proposed in \cite{demand}.
Note that we used only one channel randomly selected from whole recorded channels per each simulation.
This dataset contains various types of noise and has 288 noise files in total.
\\

To generate the noisy speeches for the training data, we utilized the 9h worth of the Libri-light and MUSAN noise datasets.
Specifically, we randomly added one or two noise files of MUSAN to the speech file of Libri-light 9h, where the number of noise files, i.e., one or two, was randomly decided with a 50\% probability.
If two noise files were selected, they were mixed as a simple linear sum in the time domain.
When summing the target speech and noise signals, the SNR followed a Gaussian with a mean of 12 dB and variance of 8 dB.
Similarly, to generate the noisy speeches for the inference and evaluation, we utilized the remaining 1h of Libri-light and the DEMAND.
The speech and noise for the evaluation data were summed in the same manner as the training data, except that the SNR parameters followed Gaussian with a mean of 8 dB and a variance of 6 dB.
Note that a voice activity detection (VAD) based on the target speech power, which regards the sections as voice if they are within 15 dB of the section with the maximum power in an utterance, was applied when calculating the SNR.
We repeated the training simulations three times and the evaluation simulations once on Libri-light.
Thus, the number of noisy utterances per epoch totaled $7431\:(= 2477 \times 3)$ in the training and $286\:(= 286 \times 1)$ in the evaluation.

\subsubsection{Networks}
\label{subsec:exp_net}
\vspace{-1.5mm}
On the basis of \cite{metricgan}, we built the following DNNs for the CER estimator and SE model.
\\
\\
\textbf{CER Estimator} \\
We employed a CNN with four two dimensional (2D) convolutional layers with the following number of filters and kernel sizes: 
[75, (5, 5)], [75, (7, 7)], [75, (9, 9)], and [75, (11, 11)]. 
To handle the variable length input, i.e., speech utterance each of which has its own length, a 2D global average pooling layer was added such that the features had 75 dimensions, where 75 is the number of feature maps of the layer before GAP.
Three fully connected layers were subsequently added, the outputs of each layer with respectively 50, 10 LeakyReLU nodes, and 1 linear node.
In addition, to make the DNN-based CER estimator $D_{\mbox{\tiny{cer}}}$ a smooth function,
all layers were constrained to be 1-Lipschitz continuous by spectral normalization \cite{spectral_normalization}.
\\
\\
\textbf{SE model} \\
We employed bidirectional long short-term memory (BLSTM) with two bidirectional LSTM layers, each with 200 nodes, followed by two fully connected layers, the outputs of each layer with respectively 300 LeakyReLU and 257 sigmoid nodes for the mask estimation.
\\

In addition, we employed a pre-trained public AM, i.e., Wav2Letter proposed in \cite{wav2letter}, as the black-box ASR system.
Wav2Letter was trained by using noisy speeches, and the trained model is available at GitHub\footnote{\url{https://github.com/flashlight/wav2letter}}.
We actually ran Wav2Letter per each iteration in order to make the ground truth of CER values for training our CER predictor.

\vspace{-1.5mm}
\subsection{Results}
\label{subsec:results}
\vspace{-1.5mm}
In our experiments, we trained the MetricGAN \cite{metricgan} by using the CER as a target metric in our experiments, and denoted it as ``Baseline'' in the following discussions.
To evaluate the performances, we used seven metrics: PESQ, CBAK, COVL, CSIG, Segmental SNR (SegSNR), STOI and CER \cite{csig, stoi, pesq}.
Note that how to calculate the CER followed Eq.~\eqref{eq:cer}.
\begin{table}[tb]
\centering
\caption{
    Experimental results.
    Note that all SE models have the same network architecture and number of parameters.
    From 2nd to 4th rows, our ideas regarding data preparation are added to the baseline one by one (ablation study).
}
\vspace{-2.0mm}
\resizebox{\linewidth}{!}{
\begin{tabular}{ c | c c c c c c c}
	\hline
    \textbf{Method} 
                 & \textbf{CSIG}$\uparrow$ & \textbf{CBAL}$\uparrow$ & \textbf{COVL}$\uparrow$ & \textbf{PESQ}$\uparrow$ & \textbf{STOI}$\uparrow$ & \textbf{SegSNR}$\uparrow$ & \textbf{CER}$\downarrow$ \\ \hline \hline
Input (not enhanced) & 3.19 & 2.29 & 2.46 & 2.77 & 0.904 & 1.17 & 0.110 \\ \hline
Baseline: MetricGAN \cite{metricgan}, & \multirow{2}{*}{3.19} & \multirow{2}{*}{2.55} & \multirow{2}{*}{2.53} & \multirow{2}{*}{2.81} & \multirow{2}{*}{0.884} & \multirow{2}{*}{3.61} & \multirow{2}{*}{0.134} \\
using CER as criteria & & & & & & & \\ \hdashline
+ Normalization & \multirow{2}{*}{2.73} & \multirow{2}{*}{2.17} & \multirow{2}{*}{2.10} & \multirow{2}{*}{2.70} & \multirow{2}{*}{0.872} & \multirow{2}{*}{1.56} & \multirow{2}{*}{0.117} \\
(Sec.~\textit{\ref{subsec:data_normalize}}) &  &  &  &  &  &  &  \\ \hdashline
+ SpecAugmentation: & \multirow{2}{*}{3.20} & \multirow{2}{*}{2.32} & \multirow{2}{*}{2.50} & \multirow{2}{*}{2.92} & \multirow{2}{*}{0.908} & \multirow{2}{*}{1.39} & \multirow{2}{*}{\textbf{0.100}} \\
\textbf{CER-centric SE} (Sec.~\textit{\ref{subsec:data_specaug}}) & & & & & & \\ \hline
Conventional SE & \textbf{3.45} & \textbf{2.71} & \textbf{2.77} & \textbf{3.02} & \textbf{0.911} & \textbf{4.86} & 0.111 \\ \hline
\end{tabular}
\label{tb:ablation}
}
\vspace{-2.5mm}
\end{table}
%

First, we conducted an ablation study to confirm the validity of our data preparation described in Sec.~\textbf{\ref{subsec:proposed_data}}.
As shown in the middle block of Table~\ref{tb:ablation}, 
the results of ``Baseline'' showed the better performances in terms of the perceptual quality and the degree of noise level suppression (see the composite measures and segmental SNR), but its score of CER was worse by comparing with the corresponding those of input signals.
Namely, the baseline could not improve the CER, although it was set as the target metric.
On the other hand, the methods including our ideas showed the better scores of CER than the baseline.
By adding our ideas one by one, it is confirmed that the score of CER was gradually improved. 
In particular, only the bottom of the middle block in Table~\ref{tb:ablation}, i.e., proposed method, succeeded to outperform the score of CER of input signals (see 1st and 4th rows).
From these results, we argue that our two considerations, i.e., (a) the normalization described in Sec.~\textit{\ref{subsec:data_normalize}} and (b) the SpecAugmentation described in Sec.~\textit{\ref{subsec:data_specaug}}, enabled the DNN-based SE model to improve the performance of the ASR system.
\begin{figure}[!t]
  \centering
  \subfloat[Input (not enhanced)]{
    \includegraphics[width=4.2cm, height=1.7cm]{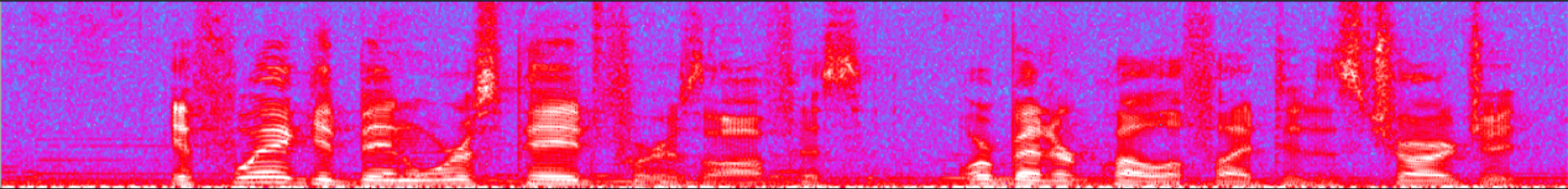}
    \label{fig:input_spec}}
  \subfloat[Clean]{
    \includegraphics[width=4.2cm, height=1.7cm]{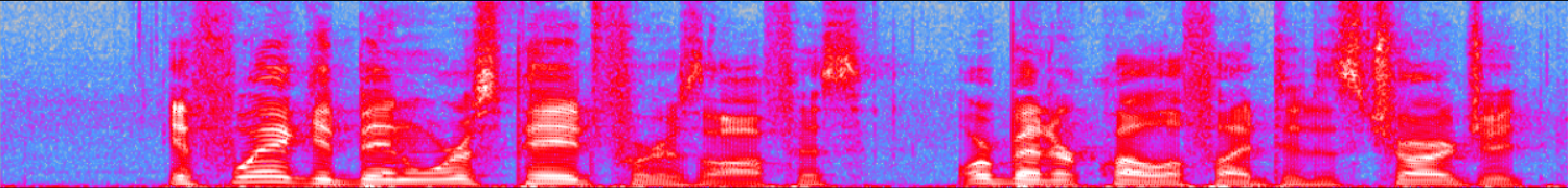}
    \label{fig:clean_spec}}

  \subfloat[Conventional SE]{
    \includegraphics[width=4.2cm, height=1.7cm]{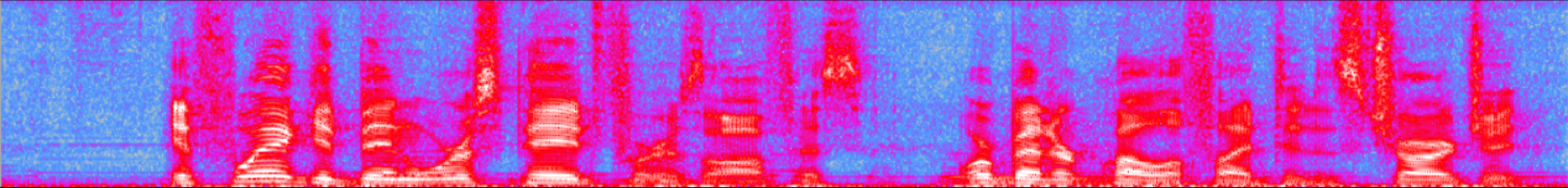}
    \label{fig:MSE_spec}}
  \subfloat[CER-centric SE (\textbf{proposed})]{
    \includegraphics[width=4.2cm, height=1.7cm]{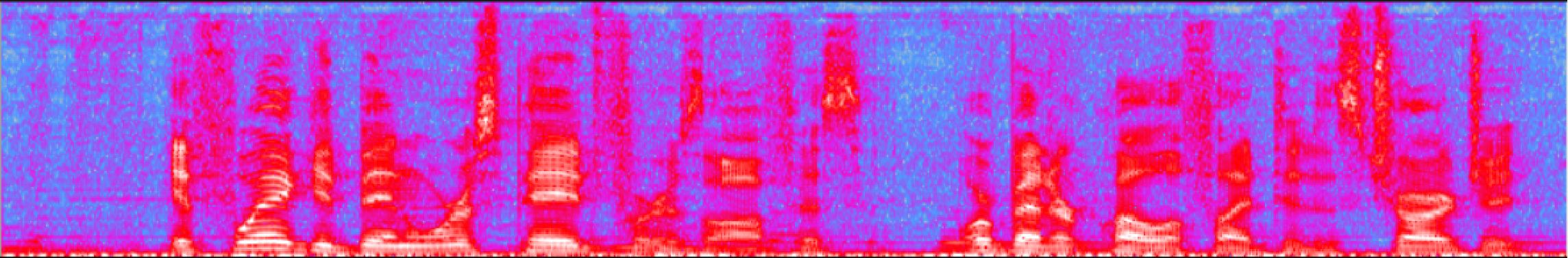}
    \label{fig:proposed_spec}}
  \caption{Spectrograms of noisy input, clean target and processed by conventional and proposed method.}
  \label{fig:comp_spec}
\end{figure}

Next, we compared the CER-centric SE model with the conventional one, i.e., 4th and 5th rows in Table~\ref{tb:ablation}.
Note that there was only one different aspect in that it was trained by applying standard MSE to the differences between the output and ground-truth clean spectrograms instead of our CER-centric training scheme.
As shown in Table~\ref{tb:ablation}, the conventional SE model showed improvements in all metrics except the score of CER compared to those of input noisy signals.
On the other hand, our method, i.e., the CER-centric SE, certainly improved the score of CER only although all metrics except for CER were inferior to those of the conventional SE model.
Actually, much more noise obviously remained in our result compared with the conventional SE model's one as shown in Fig.~\ref{fig:comp_spec}. 
From these results, we can conclude that aiming at clean speech does not always improve CER.  
In other words, the effective degree of noise level suppression that improves CER depends on which AM is in the ASR system.
Therefore, these experimental results show that conventional SE models aiming to bring the input noisy signal close to the clean one do not contribute much to improving CER.
On the other hand, our method could improve CER by 8.8\% relative keeping the appropriate degree of noise level suppression which is suitable for the ASR system.

\vspace{-2mm}
\section{Conclusion}
\label{sec:conclusion}
\vspace{-2mm}
In this paper, we proposed a new scheme of DNN-based SE aiming to minimize CER, i.e., a CER-centric training scheme for the existing SE models.
To optimize a DNN-based SE model in terms of a non-differentiable CER, two DNNs, our method alternately trains two DNNs: one for speech processing and one for mimicking the CERs calculated by using the text results through the black-box AM.
In summary, our method makes the CER function differentiable via the built CER estimator.
By using it, the existing DNN-based SE methods can be theoretically optimized to any ASR system without having to make changes in the inference phase.
Experimental results showed that the conventional SE model becomes a CER-centric model as a result of applying our method, and that it improves ASR performance. 

\clearpage
\bibliographystyle{IEEEbib}
\fontsize{8.5pt}{9.25pt}\selectfont{\bibliography{refs_euc}}

\end{document}